\documentclass[useAMS,usenatbib]{mn2e}
\input{epsf}

\usepackage{amssymb}
\usepackage{amsfonts}
\usepackage{epsfig}
\usepackage{psfig}

\def\gsim{ \lower .75ex \hbox{$\sim$} \llap{\raise .27ex \hbox{$>$}} }
\def\lsim{ \lower .75ex \hbox{$\sim$} \llap{\raise .27ex \hbox{$<$}} }
\def\Mo{{\rm M_\odot}}

\title[Globular clusters, satellite galaxies and stellar haloes from early 
dark matter peaks]
{Globular clusters, satellite galaxies and stellar haloes from early dark matter peaks}

\author[B. Moore, J. Diemand, P. Madau, M. Zemp \& J. Stadel]
{Ben Moore$^1$\thanks{moore@physik.unizh.ch.},
Juerg Diemand$^2$, Piero Madau$^2$, Marcel Zemp$^{1,3}$ \& Joachim Stadel$^1$\\
1. Institute for Theoretical Physics, University of Z\"urich
,CH-8057 Z\"urich, Switzerland\\
2. Department of Astronomy and Astrophysics, University of
California, Santa Cruz, CA 95064, USA\\
3. Institute of Astronomy, ETH Z\"urich, ETH H\@onggerberg HPF D6, CH-8093, Z\"urich, Switzerland
}

\begin{document}

\pagerange{\pageref{firstpage}--\pageref{lastpage}} \pubyear{2004}

\maketitle

\label{firstpage}  

\begin{abstract}
The Milky Way contains several distinct 
old stellar components that provide a fossil record of its formation. 
We can understand their spatial distribution and kinematics in a hierarchical 
formation scenario by associating the proto-galactic fragments envisaged 
by Searle and Zinn (1978) with the rare peaks able to cool gas 
in the cold dark matter density field collapsing at redshift $z>10$. 
We use hierarchical structure formation simulations to explore the 
kinematics and spatial distribution of these early star-forming structures in 
galaxy haloes today.
Most of the proto-galaxies rapidly merge, their stellar contents 
and dark matter becoming smoothly distributed and forming the inner Galactic halo. 
The metal-poor globular clusters and old halo stars become tracers of this early 
evolutionary phase, centrally biased and naturally reproducing the observed steep 
fall off with radius. The most outlying peaks fall in late and survive to 
the present day as satellite galaxies. The observed
radial velocity dispersion profile and 
the local radial velocity anisotropy
of Milky Way halo stars are successfully reproduced
in this model. If this epoch of structure formation coincides with a suppression
of further cooling into lower sigma peaks then we can reproduce
the rarity, kinematics and spatial distribution of satellite galaxies
as suggested by Bullock et al. (2000). Reionisation at $z=12\pm2$ 
provides a natural solution to the missing satellites problem.
Measuring the distribution of globular
clusters and halo light on scales from galaxies to clusters could be used
to constrain global versus local reionisation models. If 
reionisation occurs contemporary, our model predicts a constant
frequency of blue globulars relative to the host halo mass, except for dwarf
galaxies where the average relative frequencies become smaller.
\end{abstract}


\section{Introduction}

The Milky Way is a typical bright spiral galaxy. Its
disk of stars and gas is surrounded by an extended
halo of old stars, globular star clusters and a few
dark matter dominated old satellite galaxies. For the past 30 years
two competing scenarios for the origin of galaxies and their stellar components
have driven much observational and theoretical research. Eggen, Lynden-Bell and
Sandage (1962) proposed a monolithic collapse of the Galaxy whilst Searle
and Zinn (1978) advocated accretion of numerous proto-galactic fragments.

Enormous progress has been made in understanding the structure and origin of
the Milky Way, as well as defining a standard cosmological model for structure
formation that provides us with a framework within which to understand our 
origins \cite{peebles82,jbhf02}.
Hierarchical growth of galaxies is a key expectation within a Universe whose mass
is dominated by a dark and nearly cold particle (CDM), yet evidence for an evolving hierarchy
of merging events can be hard to find, since much of this activity took place
over 10 billion years ago. The origin of the luminous Galaxy depends on the
complex assembly of its $\sim 10^{12}M_\odot$ 
dark halo that extends beyond $200$ kpc, and on how stars
form within the first dark matter structures massive enough to cool gas to
high densities \cite{whiterees}.

The Galactic halo contains about 100 old metal poor globular clusters 
(i.e. Forbes et al. 2000) each
containing up to $10^6$ stars. Their spatial distribution
falls off as $r^{-3.5}$ at large radii and half the globulars lie within
5 kpc from the centre of the Galaxy \cite{strader04}. There is no evidence for dark
matter within the globular clusters today \cite{peebles84,moore96}.
The old stellar halo population has a similar spatial distribution and a total 
luminosity of $10^8-10^9 L_\odot$ \cite{majewski00,ivezic00}. The stellar populations,
ages and metallicities of these components are very similar \cite{jbhf02}. 

Also orbiting the Galaxy are several tiny spheroidal satellite galaxies, each containing
an old population of stars, some showing evidence for more recent star-formation
indicating that they can hold on to gas for a Hubble time
\cite{gallagher94,grebel03}. Half of the dwarf satellites lie within 85 kpc, have luminosities
in the range $10^6 - 10^{8} L_\odot$ 
and are surrounded by dark haloes at least 50-200 times as massive as their baryonic 
components \cite{mateo98}. Cold dark matter models have had a notoriously hard
time at reconciling the observed low number of satellites with the predicted 
steep mass function of dark haloes \cite{kauffmann93,moore99,klypin99}. 

We wish to explore the hypothesis that cold dark matter 
dominates structure formation, the haloes 
of galaxies and clusters are 
assembled via the hierarchical merging and accretion of smaller progenitors 
(e.g. Lacey and Cole 1993). This process violently causes structures to come to a new equilibrium by
redistributing energy among the collision-less mass components.
Early stars formed in these progenitors behave as a collisionless
system just like the dark matter particles in their host haloes, and they undergo the same
dynamical processes during subsequent mergers and the buildup of larger systems 
like massive galaxies or clusters.

In a recent study, Diemand et al. (2005) used cosmological N-body simulations to 
explore the distribution and kinematics in present-day CDM haloes
of dark matter particles that originally belonged to rare peaks in the matter
density field.
These properties are particularly relevant for the baryonic tracers of early CDM structures,
for example the old stellar halo which may have originated from the early
disruption of numerous dwarf proto-galaxies \cite{bullock00},
the old halo globular clusters and also giant ellipticals \cite{Gao2004}. 

Since rare, early haloes are strongly biased towards overdense regions (e.g. 
Sheth and Tormen 1999), i.e. towards the centers of larger scale fluctuations 
that have not collapsed yet, we might expect that 
the contribution at $z=0$ from the earliest branches 
of the merger tree is much more centrally concentrated than the overall halo.
Indeed, a ``non-linear'' peaks biasing has been discussed by previous authors  
\cite{Moore1998,White2000,moore01}. Diemand et al. (2005) showed 
that the present-day distribution and kinematics of material depends 
primarily on the rareness of the peaks of the primordial density fluctuation
field that the selected matter originally belonged to, i.e. when selecting 
rare density peaks above $\nu\sigma(M,z)$ [where $\sigma(M,z)$ is the linear theory
rms density fluctuations smoothed with a top-hat filter of mass $M$ at redshift $z$],
their properties today depend on $\nu$ and not on the specific 
values of selection redshift z and minimal mass M.

In the following section of this paper we discuss a model for the combined evolution of the
dark and old stellar components of the Galaxy within the framework of 
the $\Lambda$CDM hierarchical model \cite{peebles82}. 
Many previous studies have motivated and touched upon aspects of this work 
but a single formation scenario for the above components has not been 
explored in detail and compared with data 
\cite{kauffmann93,bullock00,cote00,moore01,jbhf02,benson02a,cote02,somerville2003,kravtsov04,kravtsov05}. 
We assume proto-galaxies and globular clusters form within the first rare
peaks above a critical mass threshold that can allow gas to cool and form stars
in significant numbers (typically at $z\approx 12$).
We assume that shortly after the formation of these first systems, the universe reionises,
perhaps by these first proto-galaxies, 
suppressing further formation of cosmic structure until later epochs. 
We use the N-body simulations to trace the rare peaks to $z=0$. Most of these
proto-galaxies and their globular clusters merge together to create the central
galactic region. In Section 3 we will compare the spatial distribution and 
orbital kinematics of these
tracer particles with the Galactic halo light and old metal poor
globular clusters. We will see that a small number of these first stellar systems
survive as dark matter dominated galaxies. We will compare their properties with the old 
satellites of the Galaxy in Section 4.

\section{The first stellar systems}\label{Sim}

\begin{figure}
\epsfxsize=8cm
\epsfysize=15.2cm
\epsffile{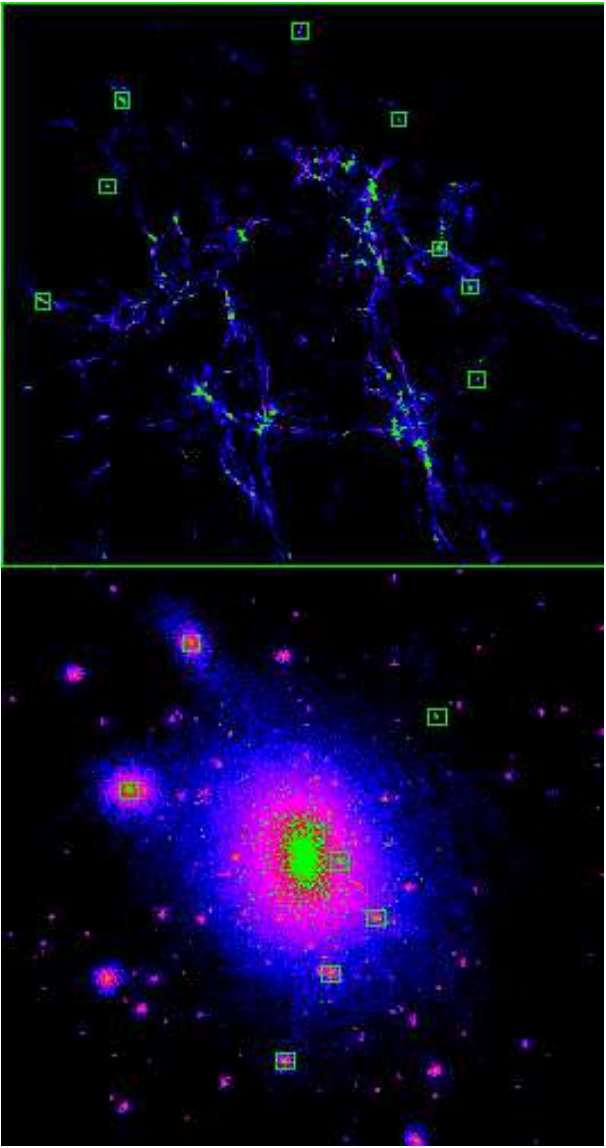}
\caption[]{
The high redshift and present day mass distribution in a region that forms a 
single galaxy in a hierarchical cold dark matter Universe.
The upper panel shows the density distribution at a redshift $z=12$ from a
region that will form a single galaxy at $z=0$ (lower panel).
The blue-pink colour scale shows the density of dark matter whilst
the green regions show the particles from proto-galaxies with virial temperature
above $10^4$ K that have collapsed at this epoch.
These peaks have masses in the range $10^8-10^{10}\,M_\odot$. 
The lower panel shows same mass distribution at $z=0$. Most of the rare peaks are located
towards the centre of the galaxy today. The squares in both panels indicate those first
objects that survive the merging process and can be associated with the visible 
satellite galaxies today orbiting within the final galactic mass halo. Most of
the subhaloes stay dark since they collapse later after reionisation has increased
the Jeans mass. 
}
\label{fig:z12}
\end{figure}

We propose that `ordinary' Population II stars and globular clusters first appeared in 
significant numbers at redshift $>12$, as the gas within protogalactic haloes 
with virial temperatures above $10^4$K (corresponding to masses comparable to 
those of present-day dwarf spheroidals) cooled rapidly due to atomic 
processes and fragmented.
It is this `second generation' of subgalactic stellar systems, aided perhaps by an earlier 
generation of metal-free (Population III) stars and by their remnant black holes, 
which generated enough ultraviolet radiation to reheat and reionize most of the hydrogen in 
the Universe by a redshift $z=12$, thus preventing further accretion of gas into the shallow 
potential wells that collapsed later. 
The impact of a high redshift UV background on structure formation has been invoked 
by several authors \cite{haardt96,bullock00,moore01,barkana01,tully02,benson02a} 
to explain the flattening of the faint end of 
the luminosity function and the missing satellites problem within our Local Group.
Here we use high resolution numerical simulations that
follow the full non-linear hierarchical growth of galaxy mass haloes to 
explore the consequences and predictions of this scenario.

Dark matter structures will collapse at different times, depending on their mass, but also
on the underlying larger scale fluctuations. At any epoch, the distribution of masses
of collapsed haloes is a steep power law towards low masses with $n(m)\propto m^{-2}$.
To make quantitative predictions we calculate the non-linear evolution of the matter 
distribution within
a large region of a $\Lambda$CDM Universe. The entire well resolved region is about 10 comoving
megaparsecs across and contains 61 million dark matter particles of mass $5.85\times 10^{5}M_\odot$
and force resolution of 0.27 kpc. 
This region is embedded within a larger 90 Mpc cube that is simulated at lower resolution
such that the large scale tidal field is represented.  Figure 1 shows the high-redshift and present-day
mass distribution of a single galaxy mass halo taken from this large volume. 
The rare peaks collapsing at high redshift that have had sufficient time to cool gas and form
stars, can be identified, followed and traced to the present day.
Because small fluctuations are embedded within a globally larger perturbation, the small
rarer peaks that collapse first are closer to the centre of the final potential and they
preserve their locality in the present day galaxy. The strong correlation between initial and
final position results in a system where the oldest and rarest peaks are spatially more
concentrated than less rare peaks. The present day spatial clustering of the material
that was in collapsed structures at a higher redshift only depends 
on the rarity of these peaks \cite{diemand05}.

Our simulation contains several well resolved galactic mass haloes which we use to trace the evolution
of progenitor haloes that collapse at different epochs. The first metal free Population 
III stars 
form within minihaloes already collapsed by $z>25$, where gas can cool via roto-vibrational 
levels of H$_2$ and contract. Their evolution is rapid and local metal enrichment occurs 
from stellar evolution. Metal-poor Population II stars form in large numbers in haloes above 
$M_{\rm H} \approx 10^8\, [(1+z)/10]^{-3/2}\,M_\odot$ (virial temperature $10^4\,$K), 
where gas can cool efficiently and fragment via excitation of hydrogen Ly$\alpha$. At $z>12$,
these correspond to $>2.5\,\sigma$ peaks of the initial Gaussian overdensity field: most 
of this material ends up within the inner few kpc of the Galaxy. Within the $\approx 1$Mpc 
turn-around region, a few hundred such protogalaxies are assembling their stellar systems \cite{kravtsov05}. 
Typically 95\% of these first structures merge together 
within a timescale of a few Gyrs, creating the inner Galactic dark halo and its associated old 
stellar population. 

With an efficiency of turning baryons into stars and globular clusters of the order 
$f_*=10\%$ we successfully
reproduce the total luminosity of the old halo population and the old dwarf 
spheroidal satellites. 
The fraction of baryons in dark matter haloes above the atomic cooling mass 
at redshift 12 exceeds $f_c=1\%$. A normal stellar population with a Salpeter-type
initial mass function emits about 4,000 hydrogen-ionizing photons per stellar baryon.
A star formation efficiency of 10\% therefore implies the emission of $4,000\times f_*
\times f_c\sim $ a few Lyman-continuum photons per baryon in the Universe.     
This may be enough to photoionize and drive to a higher adiabat vast portions of the 
intergalactic medium, thereby quenching gas accretion and star formation in nearby 
low-mass haloes.

\begin{figure}
\epsfxsize=9cm
\epsfysize=9cm
\epsffile{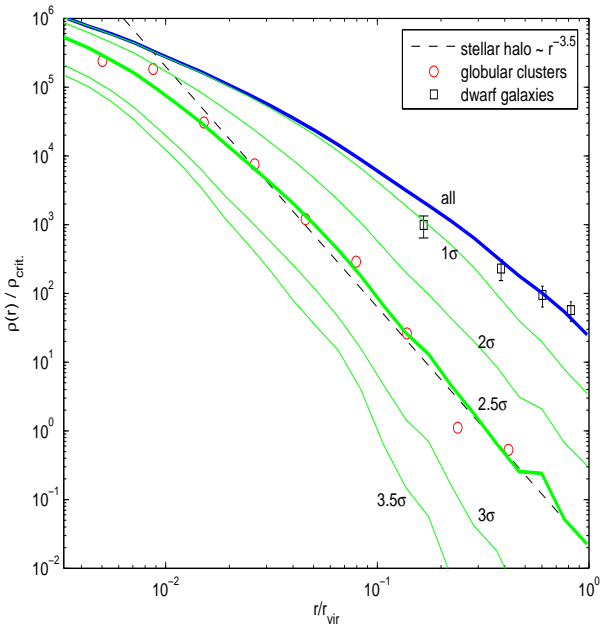}
\caption[]{
The radial distribution of old stellar systems compared with rare peaks
within a $z=0$ LCDM galaxy. The thick blue curve is
the total mass distribution today. The labeled green curves
show the present day distribution of material that collapsed into $1, 2, 2.5, 3$ and $3.5\sigma$
peaks at a redshift $z=12$. The circles show the observed spatial distribution of
the Milky Way's old metal poor globular cluster system. The dashed line indicates
a power law $\rho(r)\propto r^{-3.5}$ which represents the old halo stellar
population. The squares show the radial distribution of surviving $2.5\sigma$ peaks
which are slightly more extended than the overall NFW like mass distribution,
in good agreement with the observed spatial distribution of the Milky Way's satellites.
} 
\label{fig:z0}
\end{figure}

\section{Connection to globular clusters and halo stars}


The globular clusters that were once within the merging proto-galaxies are so dense that they 
survive intact and will orbit freely within the Galaxy. 
The surviving proto-galaxies may be the precursors of the old satellite galaxies, 
some of which host old globular clusters such as Fornax, whose morphology and 
stellar populations are determined by ongoing
gravitational and hydrodynamical interactions with the Milky Way (e.g. Mayer et al. 2005).

Recent papers have attempted to address the origin of the spatial distribution
of globular clusters (e.g. Parmentier and Grebel 2005, Parmentier et al. 2005).
Most compelling for this model and one of the key results in this paper, is that
we naturally reproduce the spatial clustering of each of these old components
of the galaxy. 
The radial distribution of material that formed from $>2.5\sigma$ peaks
at $z>12$ now falls off as $\rho(r)\propto r^{-3.5}$ within the Galactic halo - just as
the observed old halo stars and metal poor globular clusters (cf. Figure 2). 
Cosmological hydrodynamical simulations are also begining to attain the resolution 
to resolve the formation of the old stellar haloes of galaxies (Abadi et al. 2005).
Because of the steep fall off with radius, we note that we do not expect to find any 
isolated globular clusters beyond the virial radius of a galaxy
\footnote{The probability to have {\it one} isolated old globular cluster
outside of the virial radius of a Milky Way like galaxy is only 3\% in our model.}.

These first collapsing structures infall radially along filaments and end up 
significantly more flattened than the mean mass distribution. They also have colder
velocity distributions and their orbits are isotropic in the inner halo
and increasingly radially anisotropic in the outer part. Material from these rare
peaks has $\beta=1-(v_t^2/v_r^2) \approx 0.45$ at our position in the Milky Way, in 
remarkable agreement with the recently measured 
anisotropy and velocity dispersion of halo stars \cite{chiba00,bat05,thom05}.
Diemand et al. (2005) show that the radial distribution of rarer peaks is even more highly 
biased - thus the oldest
population III stars and their remnant black holes are found mainly within 
the inner kpc of the Galaxy, falling off with radius steeper than $r^{-4}$.

The observational evidence for tidal stellar streams from globular clusters suggests
that they are not embedded within extended dark matter structures today \cite{moore96}. This
does not preclude the possibility that the globular clusters formed deep within 
the central region of $10^8M_\odot$ dark haloes which have since merged together. 
(Massive substructure within the inner $\sim 20\%R_{virial}$ of galactic mass haloes
is tidally disrupted i.e. Gihgna et al. 1998.)
This is what we expect within our model which would leave the observed globulars
freely orbiting without any trace of the original dark matter component.
However, it is possible that the most distance halo globulars may still reside 
within their original dark matter halo. If the globular cluster is located
at the center of the CDM cusp, then observations of 
their stellar kinematics may reveal rising dispersion profiles. If the globular cluster
is orbiting within a CDM mini-halo then we would expect to see symmetric tidal streams
orbiting within the potential of the CDM substructure halo 
rather than being stripped by the Galaxy.

\section{Connection to satellite galaxies and the missing satellites problem}

The remaining $\sim 5$\% of the proto-galaxies form sufficiently far
away from the mayhem that they fall into the assembling galaxy late ($z\approx 1-2$, about one Gyr
after the formation of the inner Galaxy at $z\approx 5$). This leaves
time to enhance their $\alpha/{\rm Fe}$ element ratios from Type II 
supernovae \cite{wyse88,wyse95,pritzl2005}. 
Recent studies including chemical 
modeling of this process support this scenario (e.g. Robertson et al. 2005, Font et al. 2005).

The proto-galaxies highlighted with boxes in Figure 1 are those few systems 
that survive until the present epoch - they all form
on the outskirts of the collapsing region, ending up tracing the total mass distribution
as is also observed within the Milky Way's and M31's satellite systems.
Each of our four high resolution galaxies contains about ten of these surviving proto-galaxies 
which have a radial distribution that is slightly {\it shallower} than that of the total 
mass distribution but more concentrated than
the distribution of all surviving (or $z=0$ mass selected) subhalos 
(Figures \ref{fig:z0} and \ref{nr}).
This is consistent with the spatial distribution of surviving
satellites in the Milky Way and in other nearby galaxies
in the 2dF \cite{vdBosch2005,Sales2005}
and DEEP2 samples \citep{Coil2005} and with
galaxy groups like NGC5044 \citep{fal2005}.

\begin{figure}
\epsfxsize=9cm
\epsfysize=9cm
\epsffile{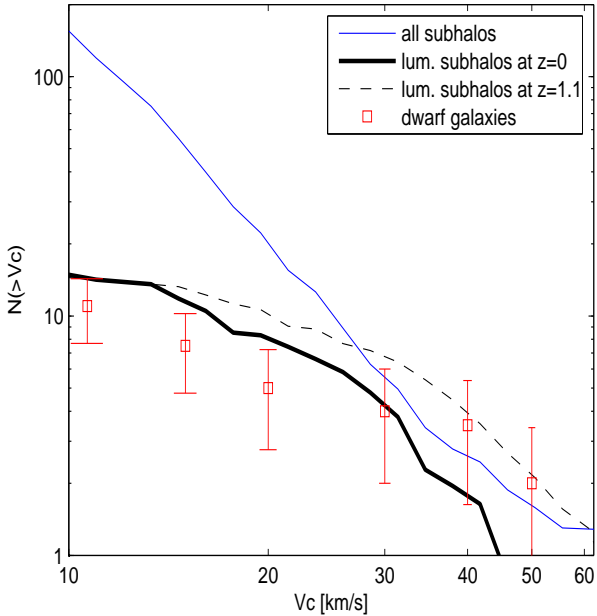}
\caption[]{
The cumulative velocity distribution function of observed Local Group satellites and predicted dark matter
substructures. The red squares show the observed distribution of circular velocities for the Local Group
satellites.
Cold dark matter models predict over an order of magnitude more dark matter substructures than are
observed within the Galactic halo (blue curve). The black solid curve 
show the cumulative velocity distribution of present day surviving substructure haloes 
that were the rare $>2.5\sigma$ peaks identified 
at $z=12$, before they entered a larger mass halo. The dashed curve shows the same objects 
but at $z=1.1$ before
they entered a larger system in the hierarchy. Their similarity shows
that little dynamical evolution has occurred for these objects.
}
\label{fig:massfn}
\end{figure}

Figure 3 shows the distribution of circular velocities of the Local Group satellites compared
with these rare proto-galaxies that survive until the present day. The Local Group circular
velocity data are the new data from Maccio' et al. (2005b) where velocity dispersions have been 
converted to peak circular velocities using the results of Kazantzidis et al. (2004).
The total number of dark matter substructures is over an order of magnitude larger than the observations.
Reionisation and photo-evaporation
must play a crucial role in suppressing star formation in less rare peaks, thus
keeping most of the low mass haloes that collapse later devoid of baryons.
The surviving population of
rare peaks had slightly higher circular velocities just before accretion (at
$z\sim 1$, dashed line in Figure 3 - see Kravtsov et al. 2004), tidal stripping inside the Galaxy halo
then reduced their masses and circular velocities and they match the
observations at $z=0$. 
Dissipation and tidally induced bar formation could 
enable satellites to survive even closer to the Galactic centre (Maccio' et al. 2005a).

Likewise to the radial distribution,  
the kinematics of the {\it surviving visible} satellite galaxies
resembles closely the one of the dark matter while the same properties
for all the surviving subhalos differ
(Figures \ref{vr} and \ref{vt}).
Within the four high resolution CDM galaxy haloes our 42 satellite
galaxies have average tangential 
and radial velocity dispersions of 0.70$\pm0.08 V_{c,{\rm max}}$ and
0.56$\pm0.07 V_{c,{\rm max}}$ respectively, i.e. $\beta = 0.26\pm0.15$ (the errors are one sigma
Poission uncertainties). These values are consistent with those of the
dark matter particles: $\sigma_{\rm tan}=0.66 V_{c,{\rm max}}$, $\sigma_{\rm 
rad}=0.55 V_{c,{\rm max}}$ and
$\beta = 0.30$; the hint of slightly larger dispersions of the satellites are
consistent with their somewhat larger radial extent. 
In the inner part our model satellite galaxies
are hotter than the dark matter background, especially in the tangential component:
Within 0.3 $r_{\rm vir}$ we find 
$\sigma_{\rm rad,GALS} / \sigma_{\rm rad,DM}=0.69 V_{c{\rm max}} / 
0.62 V_{c,{\rm max}} = 1.11$ and
$\sigma_{\rm tan,GALS} / \sigma_{\rm tan,DM}=0.95 V_{c,{\rm max}} / 0.76 
V_{c,{\rm max}} = 1.25$.
This is consistent with the observed radial velocities of Milky Way satellites. 
For the inner satellites also the tangential motions are know (with large 
uncertainties however) (e.g. Mateo 1998; Wilkinson \& Evans 1999) and 
just as in our simple model they
are larger than the typical tangential velocities of dark matter
particles in the inner halo.

The total (mostly dark) surviving subhalo population is
more extended and hotter than the dark matter while
the distribution of orbits (i.e. $\beta$) is similar \citep{Diemand2004sub}. For the 
2237 subhalos within the four galaxy haloes find
$\sigma_{\rm tan}=0.84 V_{c,{\rm max}}$, $\sigma_{\rm tan}=0.67 V_{c,{\rm max}}$ and
$\beta = 0.21$, i.e. there is a similar velocity bias relative to the dark matter
in both the radial and tangential components and therefore a similar anisotropy.
In the inner halo the differences between dark 
matter particles and subhaloes are most obvious: 
Within 0.3 $r_{\rm vir}$ we find 
$\sigma_{\rm rad,SUB} / \sigma_{\rm rad,DM}=0.91 V_{c,{\rm max}}/ 0.62 V_{c,{\rm 
max}} = 1.47$ and
$\sigma_{\rm tan,SUB} / \sigma_{\rm tan,DM}=1.21 V_{c,{\rm max}} / 0.76 
V_{c,{\rm max}} = 1.59$.  Subhalos tend to avoid
the inner halo and those who lie near the center at $z=0$ move faster
(both in the tangential and radial directions)
than the dark matter particles, i.e. these inner subhalos have large orbital energies and
spend most of their time further away from the center 
(Figures \ref{nr}, \ref{vr} and \ref{vt}, see also Diemand et al. 2004).

\begin{table}\label{haloes}
\caption{Present-day properties of the four simulated galaxy haloes.
The columns give halo name, virial mass, 
virial radius, peak circular velocity, and radius to the peak of the circular velocity curve.
The virial radius is defined to enclose a mean density of 98.6 times the
critical density.
The mass resolution is $5.85\times 10^{5}M_\odot$ and 
the force resolution (spline softening length) is 0.27 kpc.  
For comparison with the Milk Way these halos were rescaled to a peak circular velocity
of 195 km/s. In SPH simulations of the same halos we found that this
rescaling leads to a local rotation speed of 220 km/s 
after the baryonic contraction \protect\cite{maccio2005a}.
The rescaled virial radii and virial masses are given in the last two columns.}
\begin{tabular}{l | c | c | c | c | c | c }
\hline
&$M_{\rm vir}$&$r_{\rm vir}$&$V_{c,{\rm max}}$&$r_{V_{c,{\rm max}}}$&$M_{\rm MW,vir}$&$r_{\rm MW,vir}$\\
&$10^{12}\Mo$&kpc&km/s&kpc&$10^{12}\Mo$&kpc\\
 \hline
 $G0$& $1.01$ & 260 & 160 & 52.2 & 1.83 & 317\\
 $G1$& $1.12$ & 268 & 162 & 51.3 & 1.95 & 323\\
 $G2$& $2.21$ & 337 & 190 & 94.5 & 2.39 & 346\\
 $G3$& $1.54$ & 299 & 180 & 45.1 & 1.96 & 324\\
  \hline   
\end{tabular}
\end{table}

\begin{figure}
\epsfxsize=9cm
\epsfysize=9cm
\epsffile{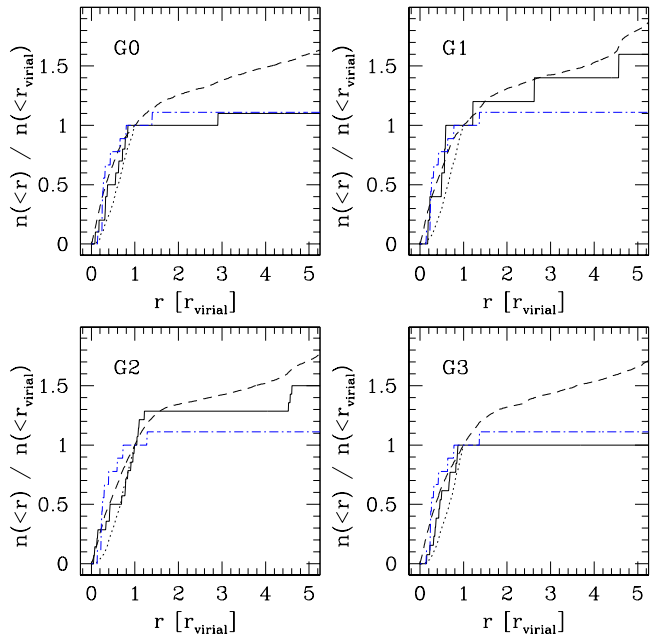}
\caption[]{Enclosed number of satellite galaxies (solid lines), all (dark and luminous) subhalos (dotted) and
dark matter particles (dashed) in four CDM galaxy halos.
The numbers are relative to the value within the virial radius. The subhalos
are only plotted out to the viral radius. For comparison with the
observed satellite galaxies around the Milky Way (dash-dotted lines) from \protect\cite{mateo98,Wilkinson99}
the simulated halos were rescaled (see Table \ref{haloes}).
The satellite galaxy distribution is more concentrated than
the one of the total surviving subhalo population but usually more extended than the dark matter particle
distribution but there are large differences from one halo to the other. Well beyond the virial
radius, the numbers of field dwarf galaxies that will host stars falls below the mean
dark matter density.}
\label{nr}
\end{figure}

\begin{figure*}
\includegraphics[width=\textwidth]{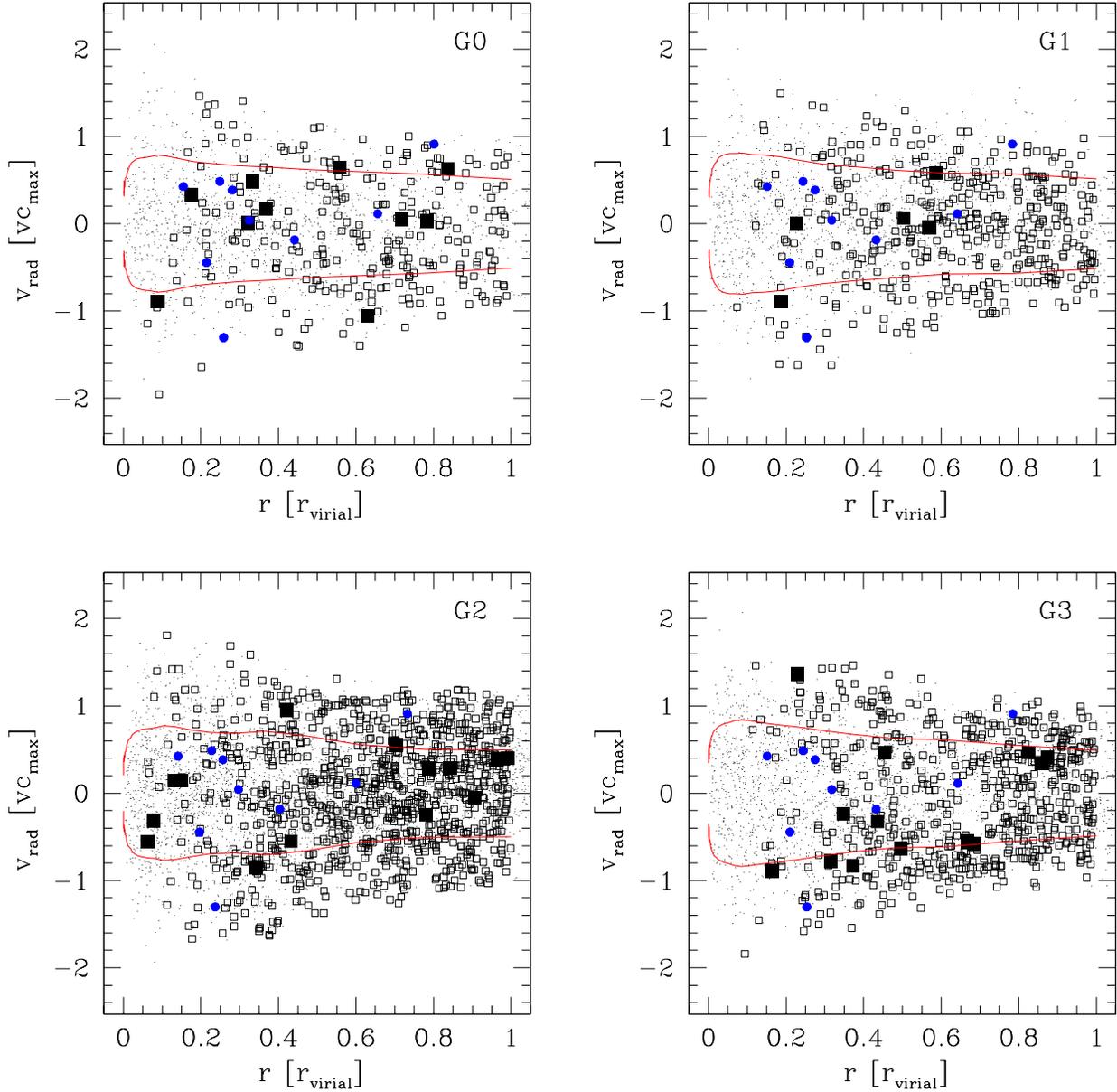}
\caption[]{Radial velocities of satellite galaxies (filled squares), 
all (dark and luminous) subhalos (open squares) and
dark matter particles (small points) in four CDM galaxy halos.
The solid lines are the radial velocity dispersion of the dark matter
plotted with positive and negative sign. All quantities are in units
of the virial radius and maximum of the circular velocity of the
host halos. For comparison with the
observed satellite galaxies around the 
Milky Way (filled circles) from \protect\cite{mateo98,Wilkinson99}
the simulated halos were rescaled (see Table \ref{haloes}).
The observed and modeled
satellite galaxies have similar radial
velocities as the dark matter particles while those of the dark subhalos
are larger, especially in the inner part. 
}
\label{vr}
\end{figure*}

\begin{figure*}
\includegraphics[width=\textwidth]{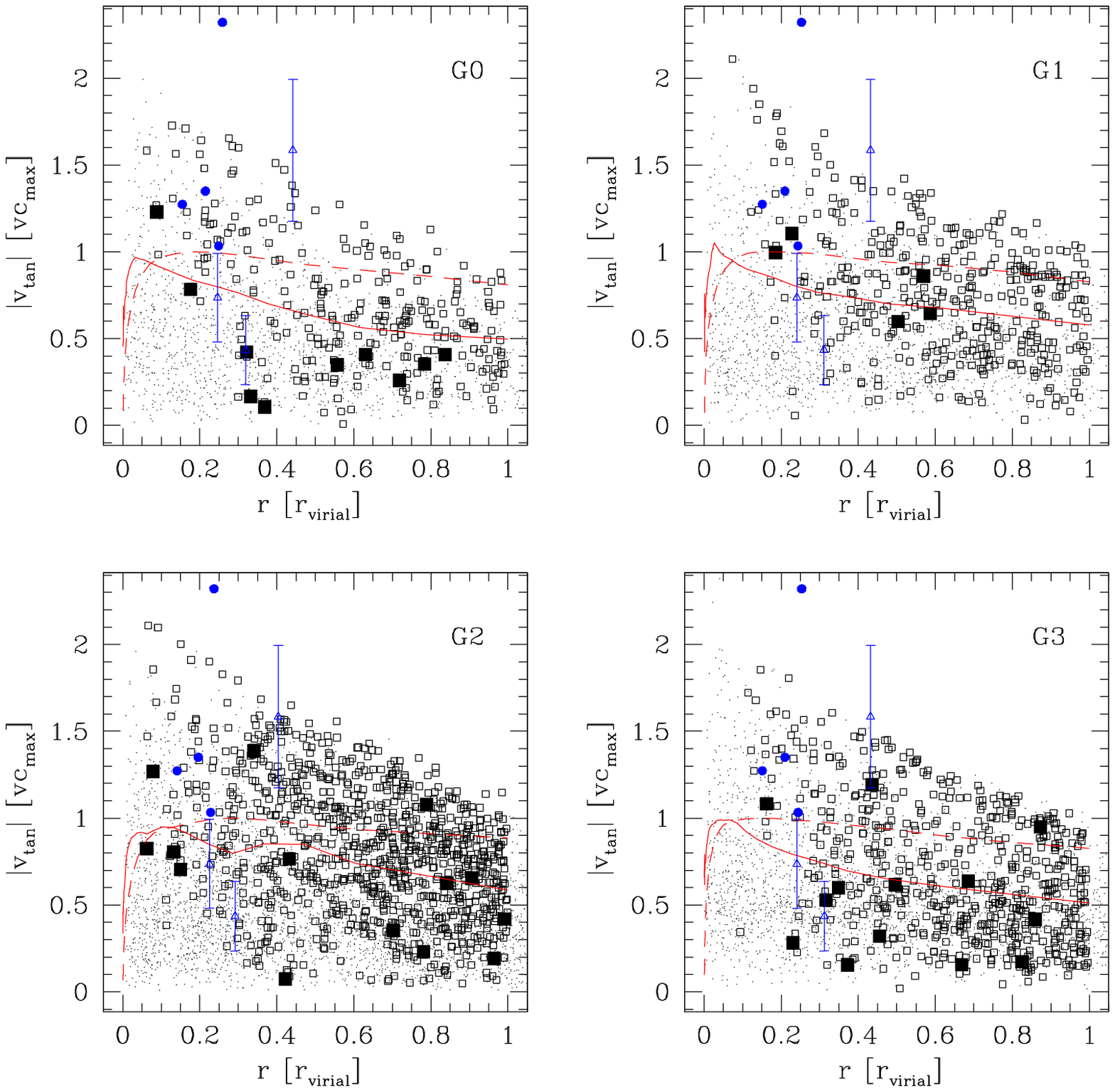}
\caption[]{Tangential velocities of satellite galaxies (filled squares), 
all (dark and luminous) subhalos (open squares) and
dark matter particles (small points) in four CDM galaxy halos.
The lines are the tangential velocity dispersion of the dark matter (solid)
and the circular velocity (dashed). The four satellite galaxies 
\protect\cite{Wilkinson99} give tangential velocities
(i.e. from inside out: LMC/SMC, Ursa Minor, Sculptor and Draco) are plotted with filled circles.
The open triangles with error bars
show HST proper motion data for (from the inside out)
Ursa Minor \protect\citep{ursaminor}, Carina \protect\citep{carina}
and Fornax \protect\citep{ursaminor}.
The units are as in Figure \ref{vr}. The observed and modeled
inner satellite galaxies (and also the dark inner subhalos)
have larger typical tangential velocities than
the dark matter particles in the same regions. 
}
\label{vt}
\end{figure*}

\section{Summary}\label{Summary and discussion}

We have a implemented a simple prescription for proto-galaxy and globular cluster
formation on to a dissipationless CDM N-body simulation. This allows us to trace the
kinematics and spatial distribution of these first stellar systems to the final
virialised dark matter halo. We can reproduce the basic properties of the Galactic
metal poor globular cluster system, old satellite galaxies and Galactic halo light.

The spatial distribution of material within a virialised dark matter structure depends
on the rarity of the peak within which the material collapses. 
This implies a degeneracy between collapse redshift and peak height. For example, 3 sigma
peaks collapsing at redshift 18 and 10 will have the same final spatial distribution within
the Galaxy. However this degeneracy can be broken since the mass and number of peaks 
are very different at each redshift. In this example at redshift 18 a galaxy mass perturbation has 700
collapsed 3 sigma halos of mass $6\times10^6M_\odot$, compared to 8 peaks of mass $4\times10^9 M_\odot$

\begin{figure}
\epsfxsize=9cm
\epsfysize=9cm
\epsffile{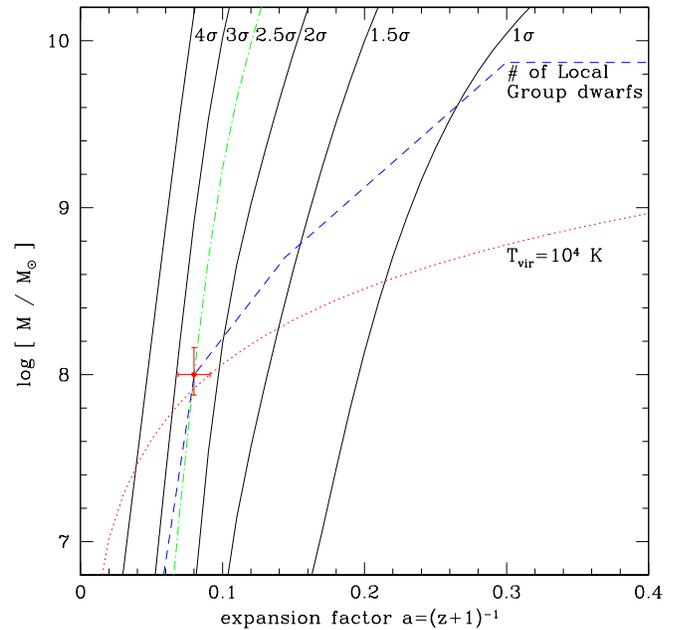}
\caption[]{Minimal halo mass for protogalaxy formation vs. time. 
The dashed line connects minimal mass/redshift pairs which produce the 
number of satellite galaxies observed in the Local Group. If efficient protogalaxy
formation is allowed below this line our model would produce too many satellites.
The best match to the spatial distribution of globular clusters and
stars comes from material that formed within peaks above 2.5 $\sigma$ (dash-dotted).
The circle with error bars indicates the latest, lowest mass
halo which is still allowed to form a protogalaxy.
The uncertainty in redshift $z=11.5^{+2.1}_{-1.5}$ is due to scatter in best fit
redshift when matching the the spatial distribution of globular clusters 
to our individual galaxy halo models at a fixed minimum mass of 10$^8 M_\odot$.
The range in minimum masses produces $N_{sat} \pm \sqrt{N_{sat}} \simeq 11 \pm 3$
luminous subhalos around an average galaxy.
The dotted line shows halos with the atomic cooling virial temperature of $10^4 K$.
Our inferred minimal mass for efficient protogalaxy formation follows the dotted line
until $z=11.5^{+2.1}_{1.5}$ and rises steeply (like the  2.5 $\sigma$
line or steeper) after this epoch.
}
\label{mvst}
\end{figure}

The best match to the spatial distribution of globular clusters and stars comes from material that
formed within peaks above 2.5 $\sigma$. We can then constrain the minimum mass/redshift pair
by requiring to match the observed number of satellite galaxies in the Local Group (Figure \ref{mvst}).
If protogalaxies form in early, low mass 2.5 $\sigma$ peaks the resulting number
of luminous satellites is larger as when they form later in heavier 2.5 $\sigma$ peaks.
We find that efficient star formation in halos above about 10$^8 M_\odot$ up to
a redshift $z=11.5^{+2.1}_{-1.5}$ matches these constraints. The scatter in redshift
is due to the different best fit redshifts found in our individual galaxy haloes.
After this epoch star formation should be suppressed in small halos
otherwise a too large number of satellites and a too
massive and too extended spheroid of population II stars is produced.
The minimum halo mass to form a protogalaxy inferred from these two constraints corresponds
to a minimal halo virial temperature of $10^4 K$ (Figure \ref{mvst}), i.e. just
the temperature needed for efficient atomic cooling.

This model is general for galaxy formation, but individual formation
histories may reveal more complexity. Soon after reionisation, infalling gas into
the already massive galactic mass halo leads to the formation of the disk and the
metal enriched population of globular clusters.
The first and second generation of stars 
forming in proto-clusters of galaxies will have a similar formation path, but 
occurring on a more rapid timescale.

\begin{figure}
\epsfxsize=9cm
\epsfysize=9cm
\epsffile{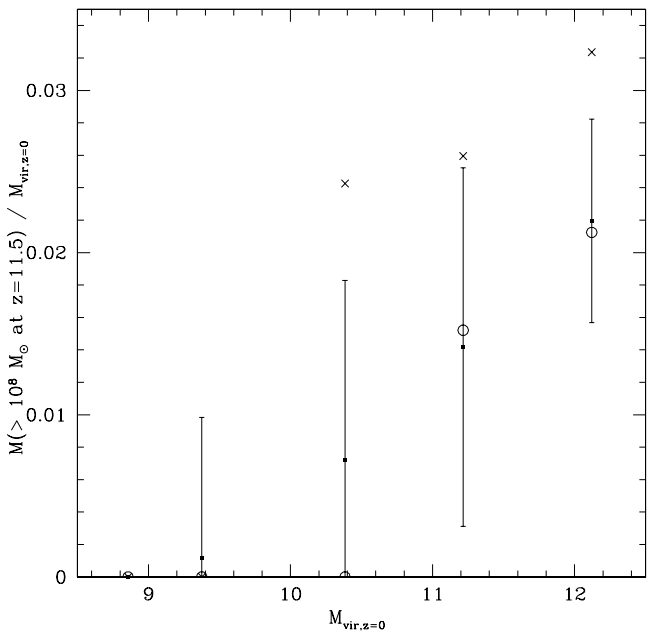}
\caption[]{Mass fraction from progenitors which are $>2.5 \sigma$ 
peaks and higher for field halos as a function of their $z=0$ virial mass. 
Filled squares with error bars are the mean values and the standard
deviations. Median (open circles) 90 percentiles (crosses) of the distributions 
are also given to
illustrate how the shape of the distribution changes for low mass hosts. 
In our simple model this mass fraction is proportional to the mass fraction of
population II stars and metal-poor globular clusters per host halo virial mass.
We have no higher mass halos in the region analyzed here but from 
Table 4 in \protect\cite{diemand05} we expect constant 
mass fractions above $10^12 M_\odot$.}
\label{fielddwarfs}
\end{figure}

We find that the mass fraction in peaks of a given $\sigma$ is independent of
the final halo mass, except that it rapidly goes to zero as the host halos become
too small to have sufficiently high $\sigma$ progenitors (see Figure \ref{fielddwarfs}
and Table 4 in \cite{diemand05}).
Therefore, if reionisation is globally coeval throughout 
the Universe the abundance of globulars normalised to
the halo mass will be roughly constant
in galaxies, groups and clusters. 
Furthermore, the radial distribution of globular clusters 
relative to the host halo scale radius will the same (see Diemand et al. 2005).
If rarer peaks reionise galaxy 
clusters earlier \cite{tully02} then
their final distribution of blue globulars will fall off more steeply 
(relative to the scale radius of the host halo) and they will be less abundant
per virial mass \cite{diemand05}. 
Observations suggest that the numbers of old globular clusters are correlated with
the luminosity of the host galaxy \cite{mclaughlin1999,Harris1991,Harris2005,rhode2005}. 
Wide field surveys
of the spatial distribution of globulars in groups and clusters may reveal 
the details of how and when reionisation occurred \cite{forbes97,puzia04}. 

\section*{Acknowledgments}
We thank Jean Brodie, Andi Burkert, Duncan Forbes and George Lake 
for useful discussions and Andrea Maccio'
for providing the corrected Local Group data for Figure \ref{fig:massfn} prior to publication.
All computations were performed on the zBox supercomputer at the University of Z\"urich.
Support for this work was provided by NASA grants NAG5-11513 and NNG04GK85G, by NSF grant 
AST-0205738 (P.M.), and by the Swiss National Science Foundation.

\bsp
\label{lastpage}
\end{document}